\documentclass[twocolumn,preprintnumbers, amsmath, amssymb]{revtex4}
\usepackage{graphics}

\def\be{\begin{equation}}
\def\ee{\end{equation}}
\def\beq{\begin{eqnarray}}
\def\eeq{\end{eqnarray}}
\def\n{\nonumber}
\def\bay{\begin{array}}
\def\eay{\end{array}}

\begin{document}
\preprint{CIRI/07-smw04}
\title{Special Relativity as a Physical Theory}

\author{Sanjay M. Wagh}
\affiliation{Central India Research Institute, \\ Post Box 606,
Laxminagar, Nagpur 440 022, India\\
E-mail:cirinag\underline{\phantom{n}}ngp@sancharnet.in}

\date{October 18, 2004}
\begin{abstract}
A modest aim of this pedagogical presentation is to analyze,
critically, certain fundamental physical concepts to illustrate
the physical principles behind the special theory of relativity
and, hence, to also illustrate the limitations of its
applicability.

\centerline{--------------------------------------------------}

\centerline{Expanded version of the talk presented during a
One-Day Seminar} \centerline{\bf 100 Years of Special Relativity}
\centerline{organized jointly by} \centerline{VMV Arts, JMT
Commerce and JJP Science College, Wardhaman Nagar, Nagpur}
\centerline{and} \centerline{Central India Research Institute,
Nagpur} \centerline{on 19th September 2004}

\centerline{--------------------------------------------------}
\end{abstract}
\maketitle

\newpage

It was in 1905 that Einstein's monumental work titled {\em Zur
Elektrodynamik bewegter K\"{o}rper\/} ({\em On the electrodynamics
of moving bodies\/}) appeared in print in {\em Annalen der Physik}
- a German journal of research in physics \cite{ein1} . This
fundamentally important research work is the reason behind our
present endeavors. It would then be also appropriate if, on this
occasion, we pause to take due cognizance of the formation and the
development of some fundamental ideas in Physics. It is therefore
my modest aim to present here a critical analysis of certain
fundamental physical concepts in relation to the special theory of
relativity and thereby to illustrate the physical principles as
well as the limitations of the applicability of this theory.

Our this critical analysis must begin with the formation and
sharpening of some fundamental physical concepts that took place
during Newton's times. It is during this Golden Era of Science
that many gifted scientists conceived, formulated and defined
sharply some of the fundamental physical concepts that we use
today.

Probably, most of us use these fundamental conceptions without
realizing the great struggles leading ultimately to their accepted
meaning. As a consequence, most of us, probably, also accept the
generalizations of these conceptions {\em uncritically}. As a
consequence, an impression may also be left with us that some
concepts are unchangeable and must always hold. Perhaps, it is
also why we, many times, rigidly adhere to them even beyond their
natural applicability.

A primary concept of Physics is that of the {\em inertia\/} of a
material body. Ordinarily, a body needs to be ``pushed'' to
produce its motion. It then displays the {\em inertia or
lethargy\/} to move.

Here, motion is to be conceived as a {\em change in the
position\/} of the material body in relation to the experimenter
or the observer. Then, speed is the rate of change of position of
a material body with time. It {\em characterizes\/} the motion of
that material body in relation to the experimenter - the observer.
Therefore, the inertia and the speed of a material body are
physical conceptions that we have derived from our ordinary, day
to day, experiences involving physical bodies.

We could then {\em postulate\/} that {\em every material body\/}
has this inertia for motion and that some ``push'' is always
required to move any material body. This is a {\em
generalization\/} that we make about {\em all\/} the conceivable
physical bodies.

We could then state that a physical or a material body continues
to remain in its state of {\em rest\/} relative to an observer or
the experimenter unless and until it is acted upon by a ``push'' -
an external agency or the {\em cause\/} of the motion. This
statement could be taken as a {\em physical law\/} - the Law of
Inertia of material bodies.

In fact, this above {\em was\/} taken to be the statement of the
Law of Inertia at the beginning of the aforementioned golden era
of science. The quantitative measure of this inertia was then
called the {\em mass}, denoted by $m$, of that material body.

From our ordinary experiences, we could then also assert that the
mass $m$ of a physical body is always a {\em positive real
number\/} since we do not, ordinarily, encounter a physical body
that ``aids'' the ``push'' producing its motion.

[Then, a body that displays properties contrary to related
ordinary experiences could be considered to possess {\em negative
mass}. However, to this date, no such material body is known to
exist.]

On the face value, there does not appear to be anything
problematic with this statement of the Law of Inertia of material
or physical bodies. However, Galileo Galilei of Pisa in Italy was
the one to sharpen \cite{dialogue} this conception on the basis of
certain experiments conducted by him.

Galileo's greatness lies not only in conceiving these experiments
but also in developing further the concept of the inertia for
motion on their basis. Before him, the emphasis was on
philosophical considerations of natural phenomena. Galileo
supported his philosophy with appropriate experimentations and
observations. It is a change of attitude from only
``philosophical'' analysis to {\bf verifiable or scientific}
analysis.

Galileo observed that when we place a material body on an inclined
plane, it has tendency to descend down it. An idea then occurred
to Galileo of placing another inclined plane next to the first one
in such a manner that the descending body would climb up this
second inclined plane.

He observed that the material body climbs the second inclined
plane up to the height above the ground from which it was released
on the first inclined plane, this when the surfaces (of contact
between the plane and that material body) were made smooth, that
is to say, as frictionless as possible. Galileo, as
extraordinarily gifted scientist as he was, then conceived a
series of careful experiments to test how far up the second
inclined plane the material body climbs when the angle of incline
of that second plane is changed.

He then noticed that as the second inclined plane is made more and
more horizontal, tangent to the surface of the Earth, the material
body travels larger and larger distance along this plane but
reaches the same height from which it was released on the first
inclined plane.

He also noticed that the material body possesses, relative to the
experimenter or the observer in the laboratory, the same speed,
say, $v$, at the bottom of the inclined plane when the height from
which it is released on the plane is the same, this irrespective
of the angle of incline of the plane. The speed $v$ was also found
to vary with the (square root of the) height from which the
material body was released. Moreover, he also noticed that this
speed $v$ is the {\em same\/} for different material bodies
(wooden, iron, glass etc.) under the same situations.

[In modern terms, the total energy $E$ is conserved in the
situation. Then, if mass of the material body is $m$, its height
above the ground is $h$ and $g$ is to denote the acceleration due
to gravity then, the material body has $E=mgh$ to begin with since
it begins from the state of rest at height $h$ above the ground.
At the bottom of the inclined plane, this potential energy is
converted into kinetic energy of the body: $mv^2/2$. Equating the
two, we obtain: $v=\sqrt{2gh}$, an expression independent of the
mass of the material body but varying as the square root of the
height $h$. Of course, this holds only in the absence of
friction.]

Galileo, then, logically argued: if the second inclined plane (of
infinite spatial extent) were made completely horizontal, {\em
ie}, exactly tangent to the surface of the Earth, then, the
material body descending down the first inclined plane would
travel an {\em infinite\/} distance rectilinearly along the second
inclined plane with the {\em uniform\/} speed $v$. By varying the
height at which the material body is released on the first
inclined plane, we would also obtain different {\em uniform\/}
speeds $v$.

The uniform rectilinear motion of a material body relative to an
experimenter therefore has no special status. A material body, as
it reaches infinite separation from the Earth, moves with {\em
uniform\/} speed even when there is no ``push'' acting on it
relative to the experimenter.

(The existence of an infinite plane is an obvious impossibility.
But, a reader of Galileo is compelled to draw the above
conclusion.)

We should then state that a physical or a material body continues
to remain in its {\em state of rest or of uniform rectilinear
motion\/} relative to an observer or the experimenter unless and
until acted upon by ``push'' - an external agency or the {\em
cause\/} of the motion. Here, we may add further that the material
body under considerations be also far removed from other physical
bodies in the universe. This is then the correct Law of Inertia of
material or physical bodies.

[The ``correctness'' of this law of inertia is as far as Galileo's
aforementioned experiments are concerned. We also note here that
{\em if some material body were found to move with the {\em
same\/} uniform speed relative to {\em all\/} the experimenters or
observers then, that body would have {\em zero\/} inertia in the
sense described above}. Nothing of Galileo's conceptions prevents
the existence of such material bodies. We will return to this
issue.]

It is our routine to state this law - Newton's First Law of Motion
- in this form. We mostly learn it as a statement of facts
without, perhaps, learning this interesting history.

However, in Galileo's times, he had to struggle \footnote{For his
purely {\em scientific views}, heresies in the eyes of the Roman
Catholic Church, Galileo had to face the Cardinal Judges of the
Inquisition Board of the Church for this heretic crime. \\
Galileo was of course not any (religious or otherwise) fanatic
person seeking martyrdom for only the sake of the beliefs he held,
particularly when he was already of an advanced age. He
felt it advisable to bend before his persecutors. \\
Galileo then had to recite, kneeling before the Cardinal Judges of
the Inquisition Board, some formula of abjuration, this
notwithstanding his advanced age and ill health. For details, see
Giorgio de Santillana (1955) {\em The Crime of Galileo\/} (The
University of Chicago Press, Chicago). \\ But, even after this
Inquisition and subsequent abjuration, Galileo was under permanent
house-arrest for the remaining eight years of his life to also
face total blindness. His scientific works still found the way to
reach out to the whole of the Europe and, ultimately, the World.\\
However, a note was to be found in Galileo's handwriting, surely
written before he turned totally blind, on the margin of his own
copy of his famous book {\em Dialogue on the Great World Systems},
for which he faced the Inquisition, as: {\em In the matter of
introducing novelties. And who can doubt that it will lead to the
worst disorders when minds created free by God are compelled to
submit slavishly to an outside will? When we are told to deny our
senses and subject them to the whim of others? When people devoid
of whatsoever competence are made judges over experts and are
granted authority to treat them as they please? These are the
novelties which are apt to bring about the ruin of commonwealths
and the subversion of the state.}} to establish it. Genuinely
speaking, it was not any easy to realize to eliminate friction
from the experimental setup. In fact, elimination of friction is
really the key element of Galileo's experiments. If we do not
eliminate friction then, a constant ``push'' is evidently needed
to make a material body move with ``uniform'' speed. Importance of
this fact must be adequately recognized by any student of physics.

Today, the word {\em inertia\/} has associated with it the meaning
of the {\em opposition\/} of a material body to a {\bf change} in
its state of rest or of uniform motion of rectilinear character.

[At this place, let us also note that one of the problems of
foremost importance for the present Physics is to ``explain'' the
{\bf origin of inertia} of material bodies.]

Clearly, the next primary conception is that of the ``push'' or
the ``external influence'' that causes the motion of a physical
body. This concept is then related to the concept of change in the
state of motion (as defined, to begin with, in the Law of Inertia
that we have stated earlier) of a material body. Then, much of
what we shall consider next will deal with the conceptions of what
this ``push'' means and also of what we really mean by a
``material'' body.

Now, let us move on to consider some more ideas. Even before
Newton, many others had realized the importance of the concept of
inertia as a fundamental property of physical bodies. Then,
efforts began to obtain the Laws of Motion using this fundamental
conception.

In this connection, Descartes then made the mathematical
construction of the Cartesian {\em coordinate system}. He realized
that, to locate a material body, we require {\em three\/} suitable
numbers in relation to another material body. In his mathematical
construction, Descartes selected a point as the origin of the
coordinate system and chose the three numbers such that the
distance of another point from the chosen origin is the Euclidean
distance: $x^2+y^2+z^2$ where $x$, $y$, $z$ denote the coordinates
of the point in relation to the origin and the three coordinate
axes which are perpendicular to each other and meeting at the
origin.

We must realize here that this above is entirely a mathematical
construction and that it has no physical implications. That is to
say, we can consider a {\em triplet}, $(x,y,z)$, of three real
numbers and consider that they form the (Euclidean) function:
$x^2+y^2+z^2$ which remains {\em invariant\/} when the triplet is
changed to, say, $(x', y', z')$.

[The transformations which keep invariant the Euclidean distance
$x^2+y^2+z^2$ are called Galilean Transformations. They form,
mathematically, a group - called the Galilean Group. This group
consists of translations of coordinates and rotations of the
cartesian coordinate system.]

Now, as a crucial step, let us represent a material body as a {\em
material point\/} in this cartesian system. Then, let $m$ be its
mass or inertia. In order to describe its motion, we must give the
values of its coordinates as functions of the time.

But, we must be really careful here in these associations of
physical character.

Firstly, we must remember that real material bodies are not
point-like but possess {\em spatial extension}. Therefore, our
this representation of a material body as a point is, in reality,
an idealization in which we simply replace that {\em extended
physical body\/} by a suitable point, let us call it the {\em
center of inertia}, to which we attribute the entire mass $m$ of
that material body.

Whether the above idealization ``correctly'' represents real
material bodies is to be checked only by ``experimentations''
involving real material bodies. That is, by verifying whether the
real material body follows the ``path of the center of inertia''
in an experimental situation.

Further, the Laws of Motion obtainable on the basis of these
considerations must ``demonstrate'' that this replacement of an
extended material body by a single material point is indeed true
in that the path of the material point (as the center of inertia)
must be obtainable from the paths of motions of other material
points (being considered to form the material body). That is, the
Laws of Motion must be consistent with the concept of the center
of inertia. This signifies the internal consistency of the
associated ideas.

Secondly, it is also necessary for us to realize that the ``time''
here must be {\em measurable\/} by a physical clock. That is to
say, the position of a material point is to be checked against the
reading of a physical clock. Thus, we have to say that when the
material point is at such and such location given by the three
cartesian space coordinates, the physical clock is {\em
simultaneously\/} showing such and such time. This {\em
simultaneity\/} is inherent in these physical associations.

If this simultaneity is not assumed then, there is no genuine
physical sense in saying that the material point has the
corresponding position. Furthermore, in the absence of any
correspondence with a physical clock, the ``time'' is simply a
label or a parameter (taking real values) and any other arbitrary
label would equally do. Time is also a parameter that is {\em
independent\/} of the cartesian space coordinates. It is,
therefore, an implicit assumption in these considerations that
such a correspondence with a physical clock exists and can be
made, as and when desired.

Then, we have the concept of the speed of the material body: the
rate of change of its (Euclidean) distance (in relation to the
origin of the coordinate system) with respect to the time.

Now, in Newton's times, collisions of material bodies were
considered to be the simplest interaction between them. Collision
changes the (initial) speeds of material bodies. Then, the above
considerations could be ``applied'' to collision of material
bodies to test their usefulness.

In other words, what we look here for are some {\em universal
laws\/} which hold in a collision of material points and {\em
verify\/} these laws in an actual collision of material bodies.

Descartes then conceived the {\em quantity of motion}:  $mv$ and
stated that this quantity is {\em conserved\/} in a collision of
material bodies. Descartes's assertion could not hold. [Notice
that speed is a {\em scalar\/} quantity. It was yet to be realized
that we need a {\em vector\/} quantity - {\em velocity\/} - that
has amplitude as well as direction. Obviously then, Descartes's
assertion could not have been true.]

Huygens, on the other hand, realized that the quantity $mv$ gets
conserved in a collision only if we assign to it a positive or
negative {\em sign\/} in a suitable manner by a convention.

So, consider the collision of two material points, initially
located on the X-axis, with motions completely along the X-axis.
Let the material point $m_1$ move away from the origin towards the
direction of the positive X-axis with initial speed $v^i_1$ with
positive sign and let the material point $m_2$ move closer to the
origin, {\em ie}, towards the direction of the negative X-axis,
with initial speed $v^i_2$ with negative sign. Let the directions
of motion of involved material points be {\em reversed\/} as a
result of their collision. Then, let their speeds after the
collision be $v^f_1$ with negative sign and $v^f_2$ with positive
sign. Then, as Huygens showed, in this case, we obtain the result:
$m_1v^i_1-m_2v^i_2=-m_1v^f_1 +m_2v^f_2$.

As we realize today, this above is a correct result. It is an
application of the Law of Conservation of Linear Momentum.

It is within this scheme of physical conceptions that Newton
developed the theoretical foundation for his famous three Laws of
Motion.

Now, Galileo did not state the Law of Inertia of Material Bodies
in the final form as we have done. It was Newton who stated it, in
his famous book, {\em The Principia}, as the First Law of
Mechanics developed by him. That is why we call it Newton's First
Law of Motion.

From the Euclidean geometry and the ``association'' of the inertia
of a material body with a point, material point, of the Euclidean
space, it is clear that the motion of a material body is
representable as a {\em curve\/} in this geometry.

A material point moving with, for example, uniform rectilinear
velocity along the X-axis is representable as the ``curve''
X-axis.

Alternatively, different types of curves of the Euclidean
geometry, straight line, circle etc.\ represent then possible
motions of a material body within this newtonian scheme.

Clearly, the ``push'' that produces the motion of a material body
is related to some appropriate property of the curve in a
Euclidean geometry. It was ``the mathematical genius'' - Newton -
who realized what this property really is.

Let us follow Newton further from here.

Firstly, we notice that the {\em displacement\/} of a material
point is definable as a {\em tangent\/} to the curve in this
Euclidean geometry. But, velocity is the displacement per unit
time and, hence, it also can be considered to be tangential to the
same curve. Similarly, Descartes's quantity of motion, now,
$m\vec{v}$, the {\em momentum\/} vector $\vec{p}$, is also
tangential to the same curve.

A change in the {\em velocity\/} vector of a material body, {\em
the acceleration}, is then a {\em vectorial\/} quantity. Then, the
``push'' must be related, within this newtonian scheme, to the
rate of change, with time, of some quantity tangential to the
curve. The push, that will, henceforth, be called the {\em force},
is then another vectorial quantity.

Almost prophetically, Newton then postulated his Second Law of
Motion that the force is equal to the rate of change of the vector
of the momentum. In terms of our usual notations:
\[ \vec{F}\;=\;\frac{d\vec{p}}{dt} \]
This equation of Newton's Second Law of Motion appears almost
prophetic because Newton could easily have chosen the force to be
proportional to the rate of change of velocity. But, in that case,
we would have
\[ \vec{F}\;=\;\frac{d\vec{p}}{dt}\;=\;\frac{d(m\vec{v})}{dt}\;=\;
m\,\vec{a}\; \equiv \;m\, \frac{d^2\vec{x}}{dt^2}\] an expression
that holds only when the mass or the inertia is a {\em constant}.

What is then the {\em source or cause\/} of this force? It is
important to recognize that, within this newtonian scheme, only
another material point can be the source of this force. A material
point ``here'' {\em acts\/} on a material point ``there'' with the
specified force. The newtonian scheme is then an {\em action at a
distance\/} framework.

We can then consider a physical body as many material points and
vectorially add the forces exerted by each one on the other.

Now, it remains to check whether these conceptions are applicable
to real material bodies in that these conceptions should be
self-consistent in the sense described earlier. Then, the center
of inertia or mass is to be obtained from a distribution of
material points and it ought to be shown that the center of
inertia follows a path that is obtainable from the paths of motion
of the material points considered to form the material body.

It is then History that Newton analyzed related conceptions for a
situation of two material points. An immediate question then
arises of the {\em difference\/} between the force exerted by the
first material point on the second material point, let us call
this force the {\em action}, and that exerted by the second
material point on the first material point, let us call this force
the {\em reaction}.

Surely, the most {\em natural\/} assumption here is the {\em
equality\/} of these two forces: action and reaction forces. But,
force is a vectorial quantity with amplitude and direction, both.

Newton then realized that for the internal consistency of this
theoretical framework it is necessary that these two forces must
be equal in amplitude but opposite in direction. Hence, his Third
Law of Motion: Forces of Action and Reaction are equal and
oppositely directed.

It is the remarkable {\em simplicity\/} of the newtonian scheme
that it is based on just these three Laws of Motion. The remaining
are just {\em deductions\/} that follow from these three basic
laws.

On the basis of only these remarkably simple three laws of motion,
it was then possible to calculate the planetary motions. It was
also possible to develop \cite{class-mech} the theory of tides,
the theory of the equilibrium configurations of rotating bodies,
the calculation of the speed of sound etc.

Primarily, one of the simplest forms of a general physical law is
to assert the conservation of some physical quantity when material
bodies participate in different physical processes. That a given
physical quantity is really subject to a conservation principle is
then to be decided only by performing experiments with material
bodies.

It was then the great triumph of the newtonian scheme that various
experiments confirmed different conservation principles of this
scheme: for example, those of the mass, energy, linear momentum,
angular momentum, etc.

This is the reason as to why the impact of the newtonian scheme
completely overshadowed the developments in Physics for the next
few centuries. That the mathematical structure of the newtonian
scheme developed by many others after Newton required no new
experiments or observations is testimony enough to say that the
physical foundation laid by Newton was completely sufficient to
support these developments.

This led the physicists of later generations to the {\em
erroneous\/} belief that the entirety of physics could be reduced
to the newtonian mechanics. In other words, they failed to
recognize clearly the limitations of the newtonian framework as we
have outlined above in brief.

It should now be clear at this stage that the limitations of
Newton's three laws of motion are really embedded within the
limitations of this newtonian scheme itself.

Of particular concern are the use in this scheme of the Cartesian
conceptions of Euclidean geometry and the associations of
properties of material bodies with the points of this space.

In this connection, we note that the coordinate transformations
which keep the Euclidean distance {\em invariant\/} are {\em very
special\/} transformations of the triplets $(x,y,z)$ of real
numbers. Newton's laws of motion are {\em invariant\/} only under
these {\em special\/} (the Galilean) transformations.

Moreover, the association of the inertia of a material body with
the points of the Euclidean space does not produce any change in
the Euclidean space. Then, the Euclidean space is an inert
background for the material bodies. This scheme then {\em explains
all phenomena as relations between objects existing in space and
time}.

A coordinate system, the construction of the coordinate axes and
clocks, must also be using the material bodies. But, precisely
this construction is left outside the scope of this newtonian
scheme. Therefore, we have to treat the construction of the
coordinate system as the one using {\em rigid\/} rods and clocks
which {\em never\/} get affected by anything happening with the
material bodies. This must be recognized as an important and
inherent drawback of the newtonian scheme.

[Let us construct a cartesian coordinate system using
``sufficiently long'' material rods, say, of iron. Ever imagined a
heavy-duty bulldozer or road-roller crossing, say, the rod
representing the X-axis, but ``not doing anything' to that rod?]

The cartesian coordinate system / space is then an {\em absolute,
meaning unchanging, coordinate system / space\/} in this scheme.

Within Newton's scheme, the {\em acceleration\/} of a material
point is, at a fundamental level, then to be referred to only this
absolute background space or the unchanging coordinate system.
Such coordinate systems are then fundamental, special, to this
description of physical systems.

Consequently, in selecting any coordinate frame accelerated with
respect to the background coordinate system, we will have to
introduce {\em fictitious\/} forces, the {\em pseudo-forces}, to
account for the phenomena involving material bodies. For example,
in selecting a {\em rotating frame\/} for describing the motion of
a material body, we have to introduce \cite{class-mech} the notion
of the {\em Coriolis Force\/} to account for the ``observed''
phenomena.

Coordinate frames in which fictitious forces do not occur are
defined to be {\em inertial frames of reference}. Then,
unambiguous physical construction of inertial frames is a problem
in Newton's theory since pseudo-forces are to be {\em defined\/}
in relation to only these frames in this scheme. This is obviously
unsatisfactory vicious circle.

Furthermore, this scheme has {\em four\/} independent coordinates:
three space coordinates and one time coordinate. This, the
newtonian, scheme is therefore {\em four-dimensional\/} of
character. However, it is, {\em not\/} any four-dimensional
distance, but, the three-dimensional Euclidean distance that is
{\em invariant\/} under the galilean transformations.

A further difficulty of the newtonian scheme is then the
following. Clearly, for the path of a material body, we use the
space coordinates as functions of the time coordinate. Then,
mathematically, the path of a material point is a ``curve'' in the
four-dimensional space of $(x,y,z,t)$.

But, it is the {\em peculiarity\/} or the {\em oddity\/} of this
newtonian scheme that no transformations of the time axis are, in
any way, involved in it. In other words, the time coordinate is
the {\em same\/} for {\em all\/} the observers. Why should this be
so? Newton's theory offers no explanation here.

Moreover, it should also be clear now that, when $m=0$, the
newtonian scheme offers no laws for the motion of a material body.
Clearly, $\vec{a}\,=\,\vec{F}/m$ and the acceleration has no
meaning for $m=0$. That is to say, the newtonian scheme cannot
describe the motion of a material body that moves with the {\em
same uniform speed\/} in relation to {\em all\/} the (inertial)
observers. This must be recognized as a limitation of the
newtonian scheme if such inertia-less material bodies existed in
reality.

Notice, now, that the newtonian scheme is based on two {\em
independent\/} considerations: first, those of the law of motion
and second, those of the law of force. Then, unless and until we
specify the force acting on a material point, Newton's Second Law
of Motion will be unable to provide us the path followed by that
material point.

It is precisely for this reason that Newton had to {\em
postulate\/} a separate Law of Gravitation - Newton's Law of
Gravitation.

Here, Newton introduced a new notion of the source properties of
material bodies. Precisely, if $M$ is the {\em source\/} or the
{\em gravitational mass\/} of one material point, $m$ is the {\em
gravitational mass\/} of another material point situated at
distance $d$ from the first body then, the gravitational force of
attraction (produced by $M$ and acting on $m$) is given by the
famous expression: \[ \vec{F}_g\;=\;-\,G\frac{mM}{d^2}\,\hat{d} \]
where $G$ is Newton's constant of gravitation and $\hat{d}$ is the
outwardly directed unit vector along the line joining the two
material points with origin of the coordinates being at the
location of the material point of gravitational mass $M$.

Then, in this Law of Gravitation, the force varies inversely with
the square of the distance separating the material points. This
must be recognized as an {\em important assumption}.

[Why not any other power of $d$? Why should the expression for the
force not contain any derivatives of the space coordinates?]

Furthermore, it is essential to distinguish between the {\em
inertial mass\/} and the {\em gravitational mass\/} of a material
point. These two are conceptions of very different physical
origins.

Now, consider that various bodies of differing inertia fall freely
under the action of Earth's gravity after being released from the
same distance above the ground. In terms of Newton's Laws of
Motion: let $m_i$ be the {\em inertial\/} mass and $m_g$ be the
{\em gravitational\/} mass of a material body. Then, from Newton's
Second Law of Motion and Newton's Law of Gravitation, we have:
\[\vec{F}\;=\;m_i\,\vec{a} \;=\;-\,G\,
\frac{m_g\,M}{r^2}\hat{r}\;\equiv\;m_g\,\vec{g}\] where $M$
denotes the {\em gravitational mass\/} of the Earth, $r$ is the
distance to the material body from the center of the Earth and
$\vec{g}$ denotes the acceleration due to Earth's gravity.

Thus, the (linear) acceleration is related to the acceleration due
to gravity as: \[ \vec{a}\;=\;\frac{m_g}{m_i}\,\vec{g}\] Now, if
the ratio of $m_g$ to $m_i$ were different for different material
bodies then, they would fall with different accelerations even
when released from the same distance above the surface of the
Earth. Galileo's experiments at the leaning tower of Pisa showed
that this is not the case.

Hence, the inertial and the gravitational mass are equal to a high
degree of accuracy for known material bodies. This is an
experimental result. But, the fact that these two quantities are
equal must be recognized as another additional {\em assumption\/}
of the newtonian scheme.

Furthermore, it also follows that the {\em origin\/} of the source
properties of material bodies like the gravitational mass is {\em
unexplainable\/} within the newtonian scheme. This is so because
the newtonian scheme treats the cause of the motion of a material
body, the force, as an external agency to be postulated or
specified {\em by hand}. Therefore, although Newton's Law of
Gravitation specifies the inverse-square behavior for the force of
gravity, this law is an assumption that is not any ``explanation''
of the phenomenon of gravitation.

Now, if this scheme is to be applicable to every material body, as
Newton's First Law of Motion asserts, then, every physical
phenomenon must be explainable as an interaction of material
points. In other words, every material body must be treatable as a
material point.

Therefore, Light must also be treatable as a material point within
this newtonian scheme. This is, precisely, the reason behind
Newton proposing the Corpuscular Theory of Light.

Then, the observation that Light propagates in a straight line is
{\em consistent\/} with this picture of Light as a material point:
a material point of Light moves in a straight line unless acted
upon by a force changing its direction of motion. The reflection
of Light from the surface of a mirror is also explainable on the
basis of collisions of material points of Light with the mirror.

However, it was known in those times that the shadow of an object
illuminated by Light has {\em two\/} distinguishable regions:
first, the dark one, called the {\em Umbra\/} and second,
relatively less darker one, called the {\em Penumbra}. Light also
{\em penetrates\/} the {\em geometrical shadow\/} region near the
edge of the object and diffracts. In Newton's scheme, this must be
because of some force acting on the material points of Light. This
force is then different for different material points of Light
since the penetration by Light in the geometric shadow occurs at
various depths behind the object.

Also, in Newton's own experiments with Light, Newton observed the
phenomenon of concentric {\em (Newton's) rings}. A bright ring is
a ring-shaped region of Light. A dark ring is another ring-shaped
region of no Light. There also are more than one such concentric
bright and dark rings. Once again, within the newtonian scheme,
this must be because of some force acting on the material points
of Light. This force is then evidently different for different
material points of Light.

It is then thinkable that an explanation of these phenomena on the
basis of some hypothetical force acting on the material points of
Light is obtainable within the newtonian scheme. Any such
explanation is, however, unsatisfactory.

The pivotal reason for this is that any explanation must be
universal of character. It is only in such a situation that the
involved explanation is also the {\em simple\/} one. This
principle of simplicity of an explanation has been the driving
impetus behind scientific theories.

In relation to the (hypothetical) force acting on the material
points of Light postulated within the newtonian scheme, we could
then ask: What causes the required behavior of this force acting
on the material points of Light? Is there some {\em universal},
{\em rational\/} explanation for this?

Evidently, no such universal, rational explanation is permissible
in Newton's scheme as any force is an {\em assumption\/} for it.
(Newton, perhaps, recognized this fact.) Furthermore, the
phenomenon of {\em polarization of Light\/} has no conceivable
explanation in Newton's scheme.

We have gone to great lengths in describing this evolution of
newtonian ideas here because the conception of an inertia of a
material body is one of the basic concepts of even the modern
physical theories. Newton's theory deals with the conception of an
inertia of a material body only in one respect: by ascribing it to
a point of the space. But, the newtonian framework does not
explain the origin of inertia of a material body. Clearly, this is
also an additional limitation of Newton's theoretical framework.

For us, of much later generations, these limitations of the
newtonian scheme may appear obvious. But, it must be kept in mind
that many of these limitations were pointed out during that Golden
Era of Science itself! For example, Descartes had pointed out the
rigid nature of the coordinate system; Newton himself was
uncomfortable with the absolute nature of the space.

It was of course recognized that properties of Light are not
explainable on the basis of the Corpuscular Theory of Light. In
particular, Huygens developed the Wave Theory of Light on the
basis of the hypothesis that Light is, for example, like a wave
propagating in a medium - {\em ether}. It was then possible to
explain the phenomena of Light. In particular, the polarization of
Light received an explanation with the wave theory.

However, all these limitations of the newtonian scheme do not
lessen the stature of either Newton as a scientist or that of
Newton's theory. The concepts created by him, by others, as well
as by those who erected the mathematical framework for these
concepts are still important, except that we now know their
limitations.

Having presented an in-depth critique of the newtonian theoretical
framework, Einstein once wrote \cite{ein2}: {\em Newton, forgive
me; you found the only way which, in your age, was just about
possible for a man of highest thought and creative power. The
concepts, which you created, are even today still guiding our
thinking in physics, although we now know that they will have to
be replaced by others farther removed from the sphere of immediate
experience, if we aim at a profounder understanding of
relationships.} Indeed, true this.

Now, let us then turn to modifications of the newtonian scheme
that are necessary to explain the phenomena displayed by Light.

Then, at the present stage, we know that Light does not follow the
newtonian laws of motion since the explanations based on these
laws are not {\em satisfactory\/} ones. Importantly, the property
of polarization of Light is not even explainable within the
newtonian framework.

But, we only have two types of material bodies left out of the
newtonian scheme - those with negative inertia and those with
vanishing inertia. Then, it is thinkable that Light is one of
these two types of material bodies. This, notwithstanding
Huygens's Wave Theory for Light, is an option open for further
exploration of ideas.

But, no property of Light indicates that it has negative inertia.
[For example, speed of Light does not increase when it collides
with the mirror, say. Here is a ``Push'' acting on Light, but
Light does not ``help'' it.] Then, in Newton's scheme, the only
option is of treating Light as a material body with vanishing
inertia.

Then, obviously with an hindsight now, we can say that Light needs
to be treated as a material body with vanishing inertia, this if
we are to follow, faithfully, the overall nature of the
(scientific) development of the physical ideas beginning with the
Golden Era of Science.

But, any material body with vanishing inertia moves with the {\em
same\/} speed for {\em all\/} the inertial observers. That is,
speed of a material body of vanishing inertia is a {\em universal
constant\/} as far as inertial observers are concerned. Then,
speed of Light (in vacuum) must be a universal constant for the
inertial observers.

Now, the Galilean transformations of coordinates are clearly
insufficient to accommodate the above universality of the speed of
Light for the inertial observers. It therefore follows that we
will have to ``extend'' these transformations to some suitable
others.

However, a lesson from Newton's theory, namely that, the laws of
motion of material objects retain their form in all inertial
frames, need not be discarded. (Or, equivalently, a lesson from
Galileo's experiments that uniform rectilinear motion of inertial
observers has no special status.) Hence, we should look here for
those transformations of the space and the time coordinates that
keep the laws of motion invariant.

We therefore arrive at the starting principles used by Einstein
for the formulation of his Special Theory of Relativity:
\begin{itemize} \item {\bf The Principle of (Special) Relativity:} The laws of
Physics (eg, of motion of material bodies) retain the same form in
all inertial frames of reference.
\item {\bf The Principle of the Constancy of the Speed of Light:} The
speed of Light (in vacuum) is a universal constant (with same
value) for all the inertial observers. \end{itemize}

On the basis of our considerations so far, it should then be
expected that these two principles (of the Special Theory of
Relativity) would be sufficient to provide us a logically
consistent framework for the theory.

In this connection, we note that the first of these two principles
is, in fact, the basis of Newton's theory and, at the present
stage of our theoretical considerations, there do not exist any
reasons to give up this characteristic of the newtonian framework.
The second of these two postulates is a direct consequence of our
assumption that Light is a material body with vanishing inertia.

The theoretical framework based on these two principles may then
be expected to provide us logically consistent description of the
motion of material bodies with vanishing and non-vanishing
(positive) inertia, both. In other words, we can then expect that
this theory would describe the motion of material bodies moving
not only with speed less than but moving also with the speed of
Light relative to inertial observers.

Now, by definition, we have \[{\rm
Speed\,of\,Light}\;=\;\frac{{\rm Light\;path}}{{\rm
Time\;interval}}\] Then, the required transformations of
coordinates will also involve {\em suitable\/} transformations of
the time coordinate when we demand the constancy of the speed of
Light for all the inertial observers. This issue then brings us to
Einstein's analysis of the simultaneity of events, an event being
a physical happening in space at some instant of time. Below, we
follow Einstein's {\em original presentation\/} of this analysis
from his 1905 paper.

At point $A$ of space, let there be a clock using which an
observer at $A$ determines the time values of events in the
immediate vicinity of $A$ by associating the positions of the
hands of the clock with these events. Similarly, at another point
$B$ of the space, let there be a clock, identical in all respects
to the clock at the point $A$, using which an observer at $B$
determines the time values of events in the immediate proximity of
$B$.

Now, it is important to recognize that we are yet to establish the
existence of a {\em common time\/} for the separated locations $A$
and $B$. Evidently, this is to be done by sending (Light) signal
from location $A$ to location $B$ and reflecting it back to $A$ so
that an observer at $A$ can compare readings of clocks at these
separate locations.

If a ray of Light starts from $A$ at time $t_{_A}$, reaches and is
reflected in the direction of $A$ at $B$ at time $t_{_B}$, and
arrives again at $A$ at time $t'_{_A}$ then, the clocks at $A$ and
$B$ {\em synchronize, show same time}, if \[
t_{_B}\;-\;t_{_A}\;=\;t'_{_A}\;-\;t_{_B}\] Having this procedure
for synchronism of clocks at different space locations, we can
then extend it to all of the space.

Again, it is important to note that this above procedure for
comparing clocks at spatially separated locations is a common,
day-to-day, experience. It is by detecting a ray of Light emitted
by an object or reflected from an object that we ``see'' that
object. The above procedure is an appropriate adaptation of this
common experience. Then, the adopted procedure of synchronization
of clocks is a {\em logical abstraction\/} derived from our this
day-to-day experience.

We therefore assume that the above definition of synchronism of
clocks is free of any contradictions and that the following are
universally valid: \begin{itemize}
\item {\bf Reflexivity of Synchronism Relation:} If the clock at
$B$ synchronizes with the clock at $A$ then, the clock at $A$
synchronizes with the clock at $B$ \item {\bf Associativity of
Synchronism Relation:} If the clock at $A$ synchronizes with the
clock at $B$ and also with the clock at $C$ then, the clocks at
$B$ and $C$ also synchronize with each other in the above
procedure \end{itemize}

Clearly, these are {\em assumptions\/} and we will have to {\em
assume\/} their consistency.

Then, the hypothesis of the constancy of the speed of Light, when
expressed in terms of these quantities, is \[
c\;=\;\frac{2\,d(A,B)} {t'_{A}\;-\;t_{_A}} \] where $c$ is a
universal constant and $d(A,B)$ is the ``distance'' separating $A$
and $B$.

Here, we can clearly recognize that the newtonian scheme assumes
an infinite speed of propagation for (Light) signals. Recall that
the distance separating points $A$ and $B$ is the {\em same\/} in
all the inertial frames of reference, {\em ie}, it is an {\em
invariant\/} of the galilean transformations. Furthermore, the
absolute nature of time in galilean transformations implies that
$t'_{_A}\,=\,t_{_A}$ in {\em all\/} the inertial frames if it
holds in one inertial frame. Therefore, an infinite speed of
propagation for signals is, in principle, allowed in Newton's
theory.

Now, chose an {\em inertial\/} frame, to be called a {\em
stationary\/} frame with all its paraphernalia of coordinate axes,
measuring rods and clocks. (Remember that the unambiguous
definition of an inertial frame is a problem in Newton's theory.
We rely on the approximate validity of Newton's laws of motion for
this above purpose.)

Moreover, consider a stationary {\em rigid\/} rod lying lengthwise
along the $X$-axis of the stationary frame; and let its length be
$\ell$ as measured by a measuring-rod which is also stationary in
the same frame of reference.

Next, imagine that this rigid rod is imparted a uniform speed $v$
along the $X$-axis of the stationary frame. Then, the length of
the moving rod can be established in the following two obvious
ways:
\begin{description} \item{(a)} The observer moves together with a measuring
rod and the rod to be measured. The length of the rod is then
obtained directly by superposing the measuring rod, in just the
same way as if all three were at rest. Clearly, this length must
be equal to $\ell$.
\item{(b)} The observer establishes the system of stationary clocks and
synchronizes them as per the adopted procedure. Then, the observer
ascertains the locations of the two ends of the rod, whose length
is to be measured, in the stationary frame at a definite time. The
distance between these two points, measured by the measuring-rod
already employed, is the length of the moving rod in the
stationary frame of reference.
\end{description}

To ascertain the length of the {\em moving rod\/} in the
stationary frame of reference, we adopt the procedure (b) as
follows.

Then, consider that at the ends $A$ and $B$ of the rod, clocks are
placed which synchronize with those of the stationary frame.
Imagine also that there is a moving observer passing each of these
clocks, and that these observers apply the method of synchronism
to both the clocks.

Let a ray of Light be emitted from $A$ at the stationary frame
time $t_{_A}$, let it be reflected at $B$ at time $t_{_B}$, and
let it reach $A$ again at time $t'_{_A}$. If ${R_{_{AB}}}$ is the
length of the moving rod measured in the stationary frame then,
from the principle of the constancy of the speed of Light, we have
\[ t_{_B}\,-\,t_{_A} \;=\;\frac{R_{_{AB}}}{c\,-\,v} \;\; {\rm
and}\;\; t'_{_A}\,-\,t_{_B} \;=\; \frac{R_{_{AB}}}{c\,+\,v}\]

Clearly, $ t_{_B}\,-\,t_{_A}\,\neq\,t'_{_A}\,-\,t_{_B}$. The
observers moving with the rod then find that the two clocks were
not synchronized even when the stationary frame had them
synchronized!

Consequently, it follows that there is no {\em absolute
significance\/} to the concept of the {\em simultaneity\/} of
events under the postulate of the constancy of the speed of Light
together with the use of Light to ``measure'' length and time.

But, the length, ${R_{_{AB}}}\neq\ell$, of a moving rod is then
different in different inertial frames in motion relative to each
other!

This, of course, is only a {\em kinematical effect\/} without any
absolute significance. That is, no ``physical'' change in the
length of the rod is implied here in the sense that the molecules
of the body get pressed together. Rather, it is only that the
observer for whom the rod is in motion must, for consistency of
the ideas, consider the length of the moving rod to be
${R_{_{AB}}}\,\neq\, \ell$ in physical statements about that rod.

Here, we already see that the Laws of Motion for material bodies
with vanishing inertia, when used as signals, have direct
implications also for the Laws of Motion for material bodies of
non-vanishing inertia. Of specific importance are the kinematic
effects as above.

In Newton's theory however, the lengths of a rod determined from
the two methods (a) and (b) are {\em precisely equal}. As we have
seen earlier, this is a consequence of the galilean
transformations used by Newton's theory. Therefore, the kinematic
effects of Einstein's Special Theory of Relativity do not have any
newtonian analogues.

Now, let us turn to our task of finding the required
transformations of coordinates consistent with the hypothesis of
the constancy of the speed of Light. It will of course be
simplified by choosing the reference frames in some convenient
manner. This can be done as follows.

In the {\em stationary\/} space, consider two systems of
coordinates, each with three rigid material lines, perpendicular
to one another and issuing from a point. [Here, we want to
emphasize the physical nature of the construction of the
coordinate system.] Let each system be provided with rigid
measuring rods, alike in all respects, and clocks, which too are
alike in all respects. Let the clocks of each of these coordinate
systems be synchronized as per the adopted procedure.

Let these coordinate systems be $(x, y, z, t)$ and $(x', y', z',
t')$. Let the primed ($K'$) and unprimed ($K$) systems {\em
coincide\/} at $t'=t=0$. Then, let the system $K'$ (along with its
entire construction of the three rigid material lines, measuring
rods and clocks) be moving with the uniform speed $v$ in the
direction of the positive $x$-axis relative to the system $K$. Let
the $y$ and $y'$ axes be parallel to each other and so also be the
case with the $z$ and $z'$ axes, when $K'$ is in motion.

Therefore, any event in the stationary space can be unambiguously
defined in place and in time by the two coordinate systems, in $K$
by $(x,y,z,t)$ and in $K'$ by $(x',y',z',t')$. Our required
coordinate transformations are then the relations connecting these
coordinate quantities.

Now, since the stationary space is homogeneous by construction,
meaning no particular coordinate location of an {\em event\/} is
preferable over any other location in it, it should be clear that
the required equations must be {\em linear}.

Furthermore, if we take $\zeta\,=\,x-vt$, a point at rest in $K'$
must have coordinate values $\zeta$, $y$, $z$, independent of time
$t'$.

From the origin of the system $K'$, let a ray of light be emitted
at time $t'_{_0}$ along the $x$-axis to $\zeta$. Let this ray of
light be reflected at $\zeta$ at time $t'_{_1}$, and arrive back
at the origin at time $t'_{_2}$. Then, by construction, we must
have
\[ \frac{1}{2}\, \left( t'_{_0}\,+\,t'_{_2} \right) \;=\;t'_{_1} \]
Now, by inserting the arguments of the function $t'$ and using the
principle of the constancy of the speed of Light: \beq \frac{1}{2}
\,\left[ t'\left(0,0,0,t\right) + t'\left(
0,0,0,t+\frac{\zeta}{c-v}+\frac{\zeta} {c+v}\right)\right] \n \\
=\, t'\left(\zeta,0,0,t+\frac{\zeta}{c-v}\right)\n \eeq

\noindent Then, for infinitesimal $\zeta$, we obtain:

\[ \frac{1}{2}\,\left( \frac{1}{c-v}+\frac{1}{c+v} \right)\,
\frac{\partial t'}{\partial t} = \frac{\partial t'}{\partial
\zeta} + \frac{1}{c-v} \, \frac{\partial t'}{\partial t} \]

\noindent Or, finally, as
\[\frac{\partial\,t'}{\partial \zeta} + \frac{v}{c^2-v^2}\,
\frac{\partial \,t'}{\partial t} = 0 \] It should be noted that we
could have chosen here any other point to be the origin of the ray
of Light to obtain the same equation.

Analogous considerations of a ray of Light for $y$ and $z$ axes
give us
\[ \frac{\partial\,t'}{\partial y}\;=\;0
\;\;\;\;\;\;\;\;\frac{\partial\,t'}{\partial z}\;=\;0 \] Since
$t'$ is a linear function, we therefore obtain
\[t'\;=\;\phi(v)\, \left(t\,-\,\frac{v} {c^2-v^2}\,\zeta \right) \]
where, for brevity, it is assumed that at the origin of system
$K'$, $t'=0$, when $t=0$ and $\phi(v)$ is an unknown function.

Now, a ray of Light is also propagated with velocity $c$ in the
moving frame $K'$. Expressing this in equations will then provide
us the required quantities $x'$, $y'$, $z'$. For example, for a
ray of Light emitted in the direction of increasing $x'$ at time
$t'=0$ at the origin of $K'$, we have \[x'\;=\;c\,t' \;=\;c\,
\phi(v)\, \left(t\,-\,\frac{v} {c^2-v^2}\,\zeta \right)\] But, the
same ray of Light moves with the velocity $c\,-\,v$ when measured
in the system $K$ so that \[ \frac{\zeta}{c\,-\,v}\;=\;t \]
Whence, we obtain \[ x'\;=\;\phi(v)\, \frac{c^2}{c^2\,-\,v^2}
\,\zeta \]

In summary, we therefore obtain \beq
\label{lorentztr} t'&\;=\;&\phi(v)\;\gamma \left(t-vx/c^2\right) \n \\
x' &\;=\;&\phi(v) \gamma \left( x-vt \right) \n \\ y' &\;=\;&
\phi(v) y \\ z' &\;=\;& \phi(v) z \n \eeq where \be \gamma
\;=\;\frac{1}{\sqrt{1-v^2/c^2}} \ee

Note that an additive constant will have to be placed on the right
hand side of each of these equations if no assumptions are made as
to the initial position of the moving frame $K'$.

Note also that these transformations must form a (mathematical)
group. Then, using some of the group properties, in particular,
the existence of an inverse transformation, it can be easily shown
that $\phi(v)=1$, very generally.

Now, we have still not proved that the hypothesis of the constancy
of the speed of Light is compatible with the hypothesis of the
equality of all the inertial frames. Then, to check the internal
consistency of our formulation, we need to show it explicitly that
any ray of Light also has the same, universally constant, speed
$c$ when measured from the moving frame.

To this end, at $t'=t=0$, let a spherical wave of Light be emitted
from the common origin of these two frames. Then, if $c$ were to
denote the (constant) speed of Light, the equation of the
spherical wavefront of Light wave in the unprimed frame would be
\[ x^2 +y^2 + z^2 = c^2 t^2 \] at any later (unprimed) time $t$.

By transforming this equation with the help of the relations
(\ref{lorentztr}), the equation of the same spherical wavefront of
Light in the primed frame is obtainable as
\[ (x')^2 +(y')^2 + (z')^2 = c^2(t')^2 \] at any later (primed) time $t'$.
Thus, the wave under consideration is also a spherical wave
propagating with speed $c$ in the moving frame.

The relations (\ref{lorentztr}), now with $\phi(v)=1$, are the
famous Lorentz transformations with $\gamma$ being called the
Lorentz factor.

These transformations were already obtained by Lorentz and
Fitzgerald, and were available in the literature before Einstein
propounded his special theory of relativity in 1905. It should
also be noted here that Einstein was not aware \cite{subtle} of
these transformations before 1905.

However, Lorentz and Fitzgerald, both, had in mind the
considerations related to the ``fictitious'' ether. Consequently,
the genuine credit of showing that these transformations are
kinematical of character is entirely that of Einstein's who also
derived these transformations on the basis of only the two
postulates mentioned earlier.

Furthermore, a special mention must also be made of Henri
Poincar\'{e} who had realized \cite{subtle} that Newton's laws
need modifications. Therefore, in a sense, ``The solution
anticipated by Poincar\'{e} was given by Einstein in his memoir of
1905 on special relativity. He (Einstein) accomplished the
revolution which Poincar\'{e} had foreseen and stated at a moment
when the development of physics seemed to lead to an impasse.''

Let us now turn to physical implications of the Lorentz
transformations. To this end, let us consider a rigid sphere of
radius $R$ at rest in the moving frame $K'$, a material body
possessing spherical shape when examined at rest, with the center
of the sphere being coincident with the origin of the coordinates
of $K'$.

Then, the equation of the surface of this spherical body moving
relative to the system $K$ with velocity $v$ is \[ (x')^2 + (y')^2
+ (z')^2 = R^2 \] The equation of this surface when expressed in
$(x, y, z)$ coordinates at time $t=0$ is \[ x^2\gamma^2 +y^2 +z^2
\;=\;R^2 \] It therefore has the shape of an ellipsoid of
revolution with the axes \[\frac{R}{\gamma},\;R,\;R \]

Therefore, to an observer in the stationary frame, the
$x$-dimension of the body appears to be {\em shortened\/} in the
ratio $1:\sqrt{1-v^2/c^2}$. For $v=c$, all moving objects, when
viewed from the ``stationary'' system, shrink into plain figures.

Clearly,  when viewed from the moving frame, the same result holds
good for bodies at rest in the stationary frame.

This is then the Lorentz-Fitzgerald contraction of a material body
in motion. It is of course only a kinematical effect. In other
words, to consistently describe the physics of these bodies an
observer in the stationary frame has to consider their shortened
dimensions while there being no objectively real contraction of
these bodies.

Furthermore, we can now consider a clock that marks the time $t$
when at rest in the stationary frame  $K$, and the time $t'$ when
at rest relative to the moving frame $K'$. What is then the rate
of this clock in the latter situation, relative to the stationary
frame?

Considering that the clock is at rest at the origin of the frame
$K'$, we have $x=vt$ and $t'$ as in (\ref{lorentztr}). Hence, \be
t'=t\sqrt{1-v^2/c^2}=t-(1-\sqrt{1-v^2/c^2})t \ee Consequently, the
time marked by the clock is slow by $1-\sqrt{1-v^2/c^2}$ seconds
per second, when viewed from the stationary frame.

Clearly, the clock in the stationary frame $K$ would also run slow
when viewed from the moving frame $K'$. Therefore, this is again
only a kinematical effect with no objectively real slowing of the
clock taking place.

Now, in Newton's theory, two given velocity vectors could be
vectorially added (as 3-vectors) to obtain the resultant velocity:
$\vec{V} \,=\, \vec{v}\,+\,\vec{w}$. This Law of Composition of
Velocities also changes under the Lorentz transformations.

In the system $K'$, let a point move as per the equations: \[
x'\;=\;\alpha t',\;\;\;\;\;y'\;=\;\beta t',\;\;\;\;z'\;=\;0\]
where $\alpha$, $\beta$ are constants. Then, using the Lorentz
transformations, the motion of the point relative to the system
$K$ is described by the equations: \beq x &\;=\;& \frac{\alpha +
v} {1+v\alpha/c^2}\,t \n \\ y &\;=\;& \frac{\sqrt{1- v^2/c^2}}{1+
v\alpha/c^2}\,\beta\,t \n \\ z' &\;=\;& 0 \n \eeq

If we now set \beq V^2 &\;=\;&\left( \frac{dx}{dt}\right)^2 \;+\;
\left( \frac{dy}{dt}\right)^2 \n \\ w^2 &\;=\;& \alpha^2 \;+\;
\beta^2 \n \\ \theta &\;=\;& \tan^{-\,1}{(\beta/\alpha)} \n \eeq
then, $\theta$ is the angle between the two velocity 3-vectors
$\vec{v}$ and $\vec{w}$.

After some simple calculations, we then obtain: \[ V = \frac{
\sqrt{v^2+w^2+2vw\cos{\theta} - (vw\sin{\theta/c^2})^2}} {1 +
vw\cos{\theta/c^2}} \] an expression for $V$ in which $v$ and $w$
obviously enter symmetrically.

Clearly, if $\vec{w}$ also has a direction of the axis of $x$
then, we have \be V = \frac{v+w}{1+vw/c^2} \label{vadd} \ee an
expression which is, most often, to be found in standard text
books \cite{class-mech}.

In this last expression, let $v=c-\kappa$ and $w=c-\lambda$, where
$\kappa$ and $\lambda$ are positive so that the $v$ and $w$ are
less than $c$. Then, \[ V \;=\;c\; \frac{2c-\kappa-\lambda}{2c-
\kappa -\lambda +\kappa\lambda/c} \;<\; c \] Clearly, in a
composition of two velocities which are less than $c$, there then
always results a velocity less than $c$.

Furthermore, it also follows that the velocity of Light cannot be
altered by composition with a velocity less than that of light:
\[ V = \frac{c+w}{1+w/c} \;=\;c \] It is also clear that such
``parallel transformations'' (all the involved velocities being in
the same direction as is the case here) form a sub-group of the
full group of Lorentz transformations.

[One may now be tempted to do something like $v=c$ and $w=c$ in
the expression (\ref{vadd}) and also obtain $V=c$. However, one
must remember that no observer or the experimenter can move with
the speed of Light. It therefore does not make any sense to
calculate the relative velocity between two particles of Light as
being equal to $c$. A student of special relativity must be
careful against some such obvious traps.]

Now, it is clear that Maxwell's electromagnetism is {\em
consistent\/} with the two postulates of the special theory of
relativity. Still, to begin with, Einstein explicitly showed the
consistency of the postulates of special relativity and Maxwell's
theory of electromagnetism.

Consider then a ray of Light (an electromagnetic wave) having {\em
energy\/} $E$ and making an angle $\phi$ with the $x$-axis of the
system $K$. Let us introduce the system $K'$ that is moving in
uniform parallel translation along the $x$-axis of the system $K$
with speed $v$. Then, when measured by the system $K'$, that ray
of Light can be shown to possess energy $E'$ given by: \[
E'\;=\;E\, \frac{1 -\frac{v}{c}\cos{\phi}}{\sqrt{1-v^2/c^2}} \] We
will use this result in the following.

Imagine now a stationary body in the frame $K$ and let its energy
be $E_o$ as referred to that frame. Let the energy of the same
body be $H_o$ relative to the system $K'$ moving in parallel
translation along the $x$-axis of $K$ with speed $v$.

Let this body now send out a ray of Light (of energy $L/2$ as
referred to the frame $K$) in a direction making an angle $\phi$
with the $x$-axis and, simultaneously, an equal quantity of Light
in the opposite direction so that the body, in the meanwhile,
remains at rest in the frame $K$. Let its energy now be $E_1$
relative to the system $K$ and $H_1$ relative to the system $K'$.

Clearly, the principle of conservation of energy must apply to
this process and, by the principle of (special) relativity, with
respect to both the inertial systems $K$ and $K'$.

Then, we have
the relations \beq E_0 &\;=\;& E_1+\frac{1}{2}L+\frac{1}{2}L \n \\
H_o &\;=\;& H_1 + \frac{L}{2}\, \frac{1 -\frac{v}{c}
\cos{\phi}}{\sqrt{1-v^2/c^2}} + \frac{L}{2}\, \frac{1
-\frac{v}{c}\cos{\phi}}{\sqrt{1-v^2/c^2}} \n \\
&\;=\;& H_1 + \frac{L}{\sqrt{1-v^2/c^2}} \n \eeq

And, on subtraction, we then obtain \beq
(H_o\,-\,E_o)&\;-\;&(H_1\,-\,E_1)\n \\ &\;=\;& L
\,\left(\frac{1}{\sqrt{1-v^2/c^2}}\;-\;1\right) \n \eeq

Now, the quantity $H$ is the energy relative to the system $K'$
and the quantity $E$ is the energy relative to the system $K$.
Hence, $H-E$ is the difference in energy as referred to two
inertial frames with the body being at rest in one of them.
Consequently, $H-E$ can differ from the kinetic energy $J$ of the
body with respect to $K'$ only by an additive constant $C$. That
is, we have $H_o-E_o=J_o+C$ and $H_1-E_1=J_1+C$ since $C$ does not
change during the emission of Light.

Therefore, we have the result \be J_o\,-\,J_1\;=\; L
\,\left(\frac{1}{\sqrt{1-v^2/c^2}}\;-\;1\right) \ee That is to
say, the kinetic energy of the body referred to the system $K'$
decreases as a result of the emission of Light and the amount of
decrease is independent of the properties of that body. It however
depends on the velocity $v$.

To second order of magnitude, we then obtain: \[J_o\,-\,J_1
\;\approx \; \frac{1}{2}\,\left(\frac{L}{c^2}\right) \,v^2 \] and
we can now conclude that {\em if a body gives off energy $E$ in
the form of electromagnetic radiation, its mass decreases by
$E/c^2$}.

This is another of celebrated results of the special theory of
relativity: the equivalence of mass and energy. Einstein obtained
this result in an independent publication \cite{ein3} again in
1905.

It also follows now that the inertia or the mass of a material
body depends upon its velocity relative to the stationary frame as
\be m\;=\;\frac{m_o}{\sqrt{1-v^2/c^2}} \label{masswithv} \ee where
$m_o$ is the mass as observed by an inertial frame in which that
material body is at rest. This is an important result in many
respects.

The reason for our adapting here Einstein's 1905 paper is the
following one. His methods are based precisely on the way we make
measurements - the way we use Light to communicate. A critical
analysis of fundamental concepts must begin with only such
considerations and then can it lead to theoretical advancement.
(History has also shown this before.) It will also be important at
a later stage of our present considerations.

In the early days of special relativity, this theory was
misunderstood to a large extent because kinematical effects
predicted by it were misconceived to be objectively real. This
misunderstanding then gave rise to many paradoxes. One of the
famous early paradoxes of special relativity is known as the
Paradox of Twins.

Imagine a pair of twins borne at the same earth-time. One of them
stays on the earth and the other travels in space for some
earth-years and returns to meet his counterpart on the earth. Will
the age of the traveller be less than that of the one who stayed
on the Earth?

It is important to recognize that this question is {\em
unanswerable\/} within the special theory of relativity since this
situation necessarily involves ``non-inertial'' frames of
reference. For example, to return to the Earth, the traveller twin
has to decelerate and accelerate in the path. But, special theory
of relativity stays silent on as to how to deal with such frames
of reference.

Now, it should be evident as to what the limitations of Einstein's
this special theory of relativity really are. Once again,
noteworthy for us are the use here of the Cartesian conceptions of
Euclidean geometry for the space and the associations of
properties of material bodies with the points of that Euclidean
space.

These above are essentially the {\em same conceptions\/} as that
of Newton's theory. Only the coordinate transformations used by
the special theory of relativity are then different.

It is here that we note the contributions of H Minkowski.
Minkowski showed that different cumbersome formulas of special
relativity can be elegantly recast into concise mathematical forms
if we treat the 3-space and the time as forming a {\em
4-dimensional spacetime}.

Consider different quadruplets $(x, y, z, t)$ of real numbers
forming a 4-dimensional (mathematical) space endowed with a metric
function: \be (\triangle x)^2+(\triangle y)^2 +(\triangle z)^2 -
c^2\,(\triangle t)^2 \;=\;ds^2 \ee where $\triangle x$, $\triangle
y$, $\triangle z$, $\triangle t$ are the infinitesimal changes in
the values of $x$, $y$, $z$, and $t$. Clearly, ``Lorentz
transformations'' of quadruplets keep this metric function {\em
form-invariant}.

Notice that the above metric function is {\em not\/} a
positive-definite one. In mathematical vocabulary, such a metric
function is called a {\em pseudo-metric\/} and the corresponding
space is called a pseudo-metric space. Then, we are considering
here a 4-dimensional pseudo-metric space.

(The 4-dimensional {\em spacetime\/} of special relativity is
therefore a pseudo-metric space.)

Let a specific point of this {\em spacetime\/} be called an {\em
event}. Notice that different events can now be {\em classified\/}
on the basis of whether the 4-distance $ds$ to that event from the
origin of the coordinates is positive, zero, or negative.

When $ds=0$, we call the trajectory a {\em null trajectory}. When
$ds\,>\,0$, we call the trajectory to be {\em spacelike\/} and
when $ds\,<\,0$, we call the trajectory to be {\em timelike}.

[We have chosen the signature of the metric function to be $(+, +,
+, -)$. We could instead have chosen it to be $(-, -, -, +)$. In
that case, these above definitions of ``timelike'' and
``spacelike'' will have to be interchanged.]

We call this 4-dimensional (pseudo-metric) space the {\em
Minkowski spacetime}. It is only a mathematical construction and
no physical implications are incorporated into it unless, of
course, we make the required physical associations.

Again, all these physical associations are {\em exactly\/} as they
were in Newton's theory. Recollect that we have, after all, only
{\em extended\/} the newtonian formalism to material bodies with
vanishing inertia and have not discarded any of the physical
associations of Newton's theory.

Of course, in extending Newton's formalism to material bodies with
vanishing inertia, we also use the fact that we ``observe''
material bodies using the electromagnetic radiation, Light, as a
material body of vanishing inertia.

Clearly, any modifications can then occur for only those newtonian
concepts which rely on the (implicit) assumption that signals can,
in principle, propagate with infinite speed. Evidently,
simultaneity of events and the absolute nature of time are two
such conceptions.

Now, when we carry out the required physical associations, it
follows that the null trajectory of the Minkowski spacetime is the
path followed by Light (as a material body with vanishing
inertia), a timelike trajectory is the path followed by a material
body moving with speed less than the speed of Light, while the
spacelike trajectory would be followed by a (hypothetical)
material body moving with speed faster than the speed of Light
relative to inertial observers.

But, it should be evident that the association of the inertia of a
material body with the points of the Minkowski spacetime does not
then produce any change in the Minkowski 4-space. Then, the
Minkowski spacetime is an inert background for the material
bodies. Clearly, even the special theory of relativity {\em
explains all phenomena as relations between objects existing in
space and time.} But, as shown by Minkowski, space and time in
special relativity are now mathematically treatable as a
4-dimensional spacetime.

Now, the construction of the coordinate axes and clocks must be
using the material bodies. But, this construction is left outside
the scope also of the special theory of relativity. Hence, the
construction of the coordinate system in special relativity is
using {\em rigid\/} rods and clocks which {\em never\/} get
affected by anything happening with other material bodies. This is
surely an important drawback of the special theory of relativity
just exactly as it was of Newton's theory.

The Minkowski spacetime is then an {\em absolute, unchanging,
4-space\/} in physical considerations of the special theory of
relativity.

Hence, the {\em acceleration\/} of a material point is, at a
fundamental level, then to be referred to only the 4-dimensional
background Minkowski spacetime or the corresponding unchanging
coordinate system. Such coordinate systems, the inertial frames of
reference, are then fundamental or special to this description of
physical systems.

Consequently, even in the special theory of relativity, if we
select any coordinate frame which is accelerated with respect to
the background coordinate system, we will have to introduce {\em
fictitious\/} forces, the {\em pseudo-forces}, to account for
various physical phenomena.

The unambiguous physical construction of the inertial frames of
reference is then a problem even in special theory of relativity.

Moreover, it should also be clear now that, when $v>c$, the
special theory of relativity offers no laws for the motion of a
material body. Clearly, Lorentz transformations have no (obvious)
meaning for $v>c$. This must be recognized as a limitation of the
special theory of relativity if such material bodies with speeds
larger than the speed of Light existed in reality. (Several
physicists have speculated \cite{subtle} about the properties of
(hypothetical) super-luminal material body, Tachyon, particle so
named by G Feinberg.)

Notice now that even the special theory of relativity is based on
two {\em independent\/} considerations: first, those of the Law of
Motion and second, those of the Law of Force. The Law of Motion is
then empty of content without a Law of Force even in special
theory of relativity. That is to say, unless and until we specify
the ``force'' acting on a material point, the Special Relativistic
Law of Motion will not be able to provide us the path followed by
that material point. But, this force is, likewise in Newton's
theory, an action at a distance due to source property of material
bodies.

Then, whatsoever is the Law of Force, it must be recognized as an
{\em assumption}. Any such law is a statement that we {\em
postulate\/} regarding the ``force'' acting between material
points. This means that, essentially, we could then always raise
the question: Why not any other form for the force acting between
material points? Clearly, as long as the law of motion is
independent of the law of force, this state of affairs persists.

As an example, consider Coulomb's law from electrostatics. If $Q$
and $q$ are to denote the quantities of {\em electric charge\/}
possessed by two material bodies separated by distance $d$ then,
the electrostatic force between them is given by the well known
expression: \[ \vec{F}_e\;=\;k\,\frac{q\,Q}{d^2}\,\hat{d} \] where
$k$ is a constant and $\hat{d}$ is a unit vector along the line
joining the two charges. This force is attractive or repulsive
depending on the relative sign of the two electric charges.

It must be recognized that this law is also an {\em assumption}.
Furthermore, the quantity of charge is the {\em source\/} property
of material body by which one charged body produces the force on
another charged body by acting on it at a distance. Then, we could
always raise the above question: Why not any other form for this
force?

Then, Newton's Law of Gravitation is also an assumption
vis-\'{a}-vis the special theory of relativity. But, by imagining
an inertial frame in which two separated material bodies are at
rest (so that the magnitude of the gravitational force between
them is given by Newton's law), by imagining another inertial
frame moving with uniform velocity $v$ relative to the first one
and by the equality of the inertial and the gravitational mass, it
is easy to see that Newton's law of gravitation is not a
Lorentz-invariant statement.

Evidently, special theory of relativity requires, therefore, an
appropriate modification of Newton's Law of Gravitation so that
the Law of Gravitation be {\em compatible\/} with the postulate of
(special) relativity stated earlier.

However, it must be borne in mind that, even with this
modification, we will not be able to circumvent the problem of the
Law of Force being an assumption in special relativity. No matter
what modification of Newton's Law of Gravitation we may propose,
the basic framework would continue to be unsatisfactory.
Therefore, special relativity offers no {\em explanation\/}
whatsoever for the phenomenon of gravitation. This is exactly as
was the case with Newton's theory.

Since the law of motion (requiring the concept of the inertia for
motion of a material body) is {\em independent\/} of the law of
force (requiring the concept of the source attribute for a
material body), the equality of the inertial and the gravitational
mass of a material body is also an explicit {\em assumption\/} of
special relativity. Again, this is exactly as it was with Newton's
theory.

Then, it immediately follows that the {\em origin\/} of the source
characteristics of material bodies, like the gravitational mass of
a material body, will  {\em not be explainable\/} in special
relativity.

Clearly, this is so because special relativity, similar to
Newton's theory, treats the cause of the motion of a material
body, the force, as an agency which is essentially independent of
the origin of the inertia of that body.

Surely, the equality of the inertial and the gravitational mass of
a material body is to be interpreted to mean that the {\em same
quality\/} of a material body manifests itself, according to
circumstances, as its inertia or as its weight (heaviness). It
should therefore be evident now that whenever the {\em cause\/} of
motion is treated as being essentially {\em independent\/} of the
{\em origin\/} of the inertia of motion of a material point, we
will not be able to explain the {\em origin\/} for the cause of
motion.

Clearly, due only to this reason, the origin of ``force'' - the
cause of motion - then remains unexplained in Newton's theory as
well as in special theory of relativity.

Clearly, conclusions of either Newton's theory or of the special
theory relativity based on the law of gravitation, {\em eg},
regarding the {\em final\/} or the {\em end-state of the collapse
of physical objects\/} such as a star, cannot be ``reliable'' in
their {\em entirety\/} since any ``Law of Gravitation'' is an {\em
assumption\/} of either of these two theories. It is important
that this issue be adequately recognized.

Still, it is true that certain experimental {\em justification
exists\/} for the laws of force (for example, Newton's law of
gravitation, Coulomb's law etc.) ``assumed'' by Newton's theory.
This is in the form of the verification of predicted planetary
motions, explanation of the tides, results of many day-to-day
laboratory experiments etc.

Similarly, an experimental justification also exists for special
relativity in the form of the verification of time dilation
effects, for example, for elementary particles moving close to the
speed of Light. It also ``explains'' results \cite{ein1, subtle}
of Michelson-Morley and other experiments.

Consequently, we can ``trust'' the predictions of these theories
to certain extent. Up to what precise extent can we trust the
predictions of these theories? Evidently, this question can be
completely answered only on the basis of the theory that explains
the origin of inertia. It should also be clear that ``correct''
answer to the question of the end-state of gravitational collapse
of a star, for example, can only be obtained from a theory that
explains the origin of inertia.

However, we may also ``verify'' the predictions of these theories
by laboratory experimentations and astronomical observations. It
is such ``verifications'' of the ``assumed'' law of gravitation
that are, most often, responsible for our falling in the trap of
assuming that the law of gravitation ``has the experimental
proof'' and, therefore, that this law is unchangeable.

Nothing can be really far from the truth than this! A critical
analysis of fundamental concepts helps us avoid such
misunderstandings by showing us the limitations of such concepts.

Another extremely important aspect is that of the {\em quantum\/}
considerations. These considerations acknowledge a fundamental
limitation in the (classical) ideas by essentially recognizing
that in making a measurement of a physical quantity, an observer
will inevitably cause an ``uncontrollable'' change to the system
being observed. This aspect is particularly important for
phenomena involving microscopic bodies.

As an example, consider the measurement of the location of an
object. From our ordinary day-to-day experiences, we know that to
locate an object, we must send a signal, a ray of Light, in the
direction of that object and must receive the signal reflected
from that object.

But, a ray of Light {\em imparts\/} momentum to the object in the
process of reflection, thereby changing the earlier (measured)
value of the (linear) momentum of that object. Consequently, there
is always to be an {\em interaction\/} of the object and the
observing agency in the process of any measurement of the physical
quantity.

Now, the issue is whether we can ``control'' this ``change'' in
any conceivable manner to whatsoever extent that we desire.

If we could then, we would always be able to simultaneously
measure, both, the location and the momentum of the object to any
desired accuracy. If, fundamentally speaking, we cannot then,
newtonian and special relativistic concepts require, obviously,
modifications.

This is of course not the issue of simply being unable to achieve
the above task in a particular experimental setup due to
experimental limitations of the equipments of measurement. Rather,
this is the issue of the {\em theoretical possibility or the
impossibility\/} of the measurement of a physical quantity to any
desired accuracy.

Here, it is vital to recognize that this issue is related to the
issue of the physical construction of the coordinate system, first
raised by Descartes \cite{ein1} during the Golden Era of Science.

In the experiment of the measurement of the location of an object,
Light, as a material body, acts as a (tiny) road-roller trying to
cross the location of the given object, say, as a part of the
coordinate axis. This road-roller must have some action on the
object, then. This was the issue raised by Descartes \cite{ein1}.
Then, the earlier issue is whether this action of the road-roller
is ``controllable'' by the experimenter so that the location can
be measured to any desired accuracy.

Now, it is the tacit {\em assumption\/} of Newton's theory as well
as of Einstein's special theory of relativity that this
aforementioned action is ``precisely controllable'' in the
experimental setup. That is why the location of a material point
is precisely determinable simultaneously with its (linear)
momentum in these theories.

But, is it really so? To be sure, we must investigate this issue
further. This was what was done by Heisenberg \cite{heisenberg} at
the beginning of the 20th century - another Golden Era of Science.
Apart from those of Galileo, Newton, Descartes and Einstein,
Heisenberg's investigation of this issue is another example of the
critical analysis of related fundamental concepts.

Then, from Heisenberg's analysis, it follows \cite{heisenberg}
that {\em simultaneously measured\/} values of the location and
the momentum of that object satisfy {\em (Heisenberg's)
indeterminacy relation\/} - the basis of the theory of the
quantum.

Heisenberg's indeterminacy relation shows that the location and
the (linear) momentum of a material body, represented as a
material point, cannot be simultaneously determined to any better
accuracy than is permitted by this relation. Surely, this
indicates limitations of the concepts of Newton's theory and those
of the special relativity.

In spite of this above being the case, special relativity is
surely an {\em advance or a step\/} in the direction of a theory
that can ``explain'' the origin of inertia and, hence,
gravitation.

In this connection, it must be borne in mind that the inertia, the
opposition of a material body to a change in its state of motion,
can be expected to depend on the state of motion, speed, of that
body. This expectation is based on the analogous situation of the
opposition to motion experienced by a person located in a crowd.
This opposition to motion of a person depends on the state of
motion, speed, of that person.

That the expectation based on the above analogy holds is then
expressed by the ``variation of (inertial) mass with velocity'' in
special relativity as we have seen earlier in (\ref{masswithv}).

In a sense, it is only this single result that is conceptually the
most important one of the achievements of special relativity. It
is of course not that other results such as the ``relativization
of time'' are any less important than this.

But, only the mass-variation with velocity indicates that special
relativity is a (right) step in the direction of a theory that
could explain the origin of inertia. Indeed, Einstein was fond of
referring to special relativity as a {\em step\/} \cite{subtle}.

The origin of inertia of material bodies may then be
``explainable'' by developing the aforementioned analogy (to a
person placed in a crowd) into a suitable new theoretical
framework.

Now, concepts of Euclidean geometry will, evidently, not be useful
to this new theoretical framework because the physical
construction of its coordinate system (coordinate axes, measuring
rods and clocks) is required to be unchangeable irrespective of
the motion of material bodies in this geometry. This is, as we
have now recognized, evidently, unsatisfactory.

Then, we must use that geometry whose (physical) construction of
space coordinates changes with the motions of material bodies.
Then, as a judicious guess based on our related considerations, it
therefore follows that we need to use, for the new theory,
geometric considerations more general than those of the Euclidean
geometry.

What kind of {\em non-Euclidean geometry\/} is this new one
required to be?

Here, we must first realize that a ``material body'' and
``geometry of space'' aught to be {\em indistinguishable}. Moving
a material body from its given ``location'' should cause changes
to the construction of the coordinate system. And, that will
change the ``geometry'' because the construction of the coordinate
system is the basis of the ``metric function'' of the geometry.

Hence, this new geometry is determined by physical or material
bodies. In turn, material bodies are also determined by this new
geometry in that ``given the (pseudo) metric function of the new
geometry'' we would know how the material bodies are ``located''
relative to each other.

Evidently, these ideas are then fundamentally different from those
of Newton's theory as well as from those of special relativity. In
these new considerations, the concept of ``force'' is abandoned
and is replaced by suitable properties of the geometry - its
transformations.

Even when the basic conceptions of the new theory are required to
be quite different, it must be borne in mind that the successes of
Newton's theory as well as those of special relativity will have
to be obtainable in the new theory. Mathematically, various
formulas of the new theory must ``reduce'' to those of special
relativity and to those of Newton's theory in approximations.

It is in this above mathematical sense that the new theory will
have to incorporate special relativity and Newton's theory, both.

Now, with due respects to Minkowski's works, there is nothing
fundamentally important about the 4-dimensional formalism in
special relativity. In other words, all the physical results,
obtained using explicitly the 4-dimensional considerations, can
also be obtained using the 3-space and the time, essentially
separately.

[As a matter of fact, Einstein's original paper(s) on the special
theory of relativity did not use any of the 4-dimensional methods
which were discovered by Minkowski only later.]

However, the {\em continuum\/} of the quadruplets of four
coordinates, $(x,y,z,t)$ with all the four coordinates varying in
continuous manner, may be expected to form the basis of the theory
of the future as well. After all, we have abstracted these {\em
four\/} quantities - three for the space and one for the time -
from our ordinary, day-to-day, experiences with material bodies.

Also, to this day, there are no indications, of any kind
whatsoever, from our ordinary experiences that, to describe the
``motions'' of material bodies and, hence, to also describe the
origin of their inertia, we need to add any other quantities to
this list of the four coordinates.

From the critical analysis of related concepts that we have
accomplished above, it has also not emerged that the extra
dimensions are any necessary to this description.

Consequently, even if there existed extra dimensions, these must
somehow be ``suppressed'' so as to be not observable at the
present level of observations. Why would this always be so?
Presently, we simply do not know why.

Then, with due respects to them, it must be said that attempts
\cite{subtle} to add {\em extra dimensions}, over and above those
of the four of space and time, have not led to any ``genuine
advance'' in our understanding of the physical world around us.

In the absence of any clear indications about the possible
extra-dimensions, it is, conservatively speaking, advisable to
restrict ourselves to only the quadruplets of four coordinates:
$(x,y,z,t)$. This is what we have followed here.

Then, on the basis of the mathematical ``beauty'' of Minkowski's
4-dimensional formulation, we may be tempted to think that any
future theory of the physical world should be based only on the
relevant 4-dimensional considerations.

This last issue is then that of being able to treat the set of
quadruplets as a 4-dimensional ``pseudo-metric'' {\em manifold\/}
when {\em general transformations\/} of quadruplets are invoked.
But, transformations of quadruplets need not be continuous, let
alone differentiable. Then, usefulness of the differentiable
manifold structure or the pseudo-metric structure becomes
questionable.

What can then substitute the differentiability (partial
differential equations) being used here? This is the issue of the
most basic mathematical formalism to describe physics.

Currently, theories of measures and dynamical systems appear basic
\cite{smw-unified}.

However, any further considerations of the new theory are
obviously beyond the scope of our present endeavors.

In conclusion, therefore, I hope to have made it clear to you in
this modest presentation as to why the special theory of
relativity is only a step in the right direction. Then, the
concepts used by special relativity are not to be treated as
unchangeable. In fact, we aught to change some of the concepts
here if we aim at a profounder understanding of the physical
phenomena.

I hope also to have made it clear to you that to advance further
from the conceptions in Newton's theory and in special relativity,
we need to ``abandon'' the concept of ``force'' as an external
cause of motion and ``replace'' it with some other satisfactory
concept.

Einstein had hoped to achieve this with his general theory of
relativity. However, he had realized \cite{ein1, ein2, subtle,
smw-unified} that his dream is not materialized in the formulation
that he had proposed for the general relativity. He then searched
for a new Unified Field Theory without success.

But, towards the end of his life, he had begun seeing beyond the
vision of others, as had always been the case \cite{subtle} with
him in the past.

\medskip
\begin{acknowledgements} I am grateful to Drs. Dilip A Deshpande,
Kishor B Ghormare, Pradeep S Muktibodh  and other organizers of
the Seminar on {\em 100 years of Special Relativity\/} for giving
me an opportunity of presenting my thoughts on this occasion.
\end{acknowledgements}



\begin{thebibliography}{99}
\bibitem{ein1} See, for example, articles in (1952) {\em The Principle of
Relativity: A collection of original papers on the special and
general theory of relativity. Notes by A Sommerfeld\/} (Dover, New
York) \\ See, also, Einstein Albert (1968) {\it Relativity: The
Special and the General Theory\/} (Methuen \& Co. Ltd, London)

\bibitem{dialogue} See, for example, Galileo Galilei (1955) {\em Dialogue on the
Great World Systems\/} (Abridged text edition {\em In the
translation of T Salusbury}, Edited, corrected, annotated and
provided with a historical introduction by Giorgio De Santillana,
The University of Chicago Press, Chicago)

\bibitem{class-mech} See, for example, Goldstein H (1950) {\it Classical Mechanics\/} (John
Wiley \& Sons, New York) \\ Kibble T W B (1970) {\it
Classical Mechanics\/} (ELBS-McGraw-Hill, London) \\
Sudarshan E C G and Mukunda N (1974) {\it Classical Dynamics - A
modern perspective}, (Wiley Interscience, New York)

\bibitem{ein2} Einstein Albert (1970) in {\it Albert Einstein: Philosopher
Scientist} (Ed. P A Schlipp, Open Court Publishing Company - The
Library of Living Philosophers, Vol VII; La Salle)

\bibitem{subtle} Pais Abraham (1982) {\it Subtle is the Lord ... The science
and the life of Albert Einstein} (Clarendon Press, Oxford)

\bibitem{ein3} Einstein Albert (1905)  {\em Does the inertia of a body depend upon its energy
content?}, (1905) {\em Annalen der Physik}, 17

\bibitem{heisenberg} Heisenberg W (1949) {\it The Physical
principles of the quantum theory\/} (Dover, New York) \\ See,
also, Bohr N (1928) {\it Nature (Suppliment Series)}, April 14,
1928, p. 580

\bibitem{smw-unified} Wagh S M (2004) {\em Heuristic approach to a
{\em natural\/} unification of the quantum theory and the general
theory of relativity} {\bf Database: physics/0409057} and
references therein. \\
See, also, Wagh S M (2004) {\it Einsteinian field theory as a
program in fundamental physics} {\bf Database: physics/0404028}
and references therein.
\end{thebibliography}
\end{document}